\documentclass[a4paper,12pt]{article}

\usepackage{amsfonts}
\usepackage{amssymb}
\usepackage{epsfig}

\def\beq{\begin{equation}}
\def\eeq{\end{equation}}
\def\bea{\begin{eqnarray}}
\def\eea{\end{eqnarray}}

\def\u1{\widehat{U(1)}}
\def\su2{\widehat{SU(2)}_1}

\def\a{\alpha}

\catcode`\@=11
\def\numberbysection{\@addtoreset{equation}{section}
        \def\theequation{\thesection.\arabic{equation}}}
\numberbysection

\begin{document}

\begin{titlepage}

\begin{center}
\hfill  \quad cond-mat/08012367 \\
\vskip 1 cm
{\LARGE \bf Effective Field Theories }
\vspace{.4cm}
{\LARGE \bf for Electrons in Crystalline Structures}

\vspace{1cm}

\vspace{.4cm}
Federico~L.~ BOTTESI ,\ \ Guillermo~R.~ZEMBA\footnote{
Fellow of CONICET, Argentina.}\\
\bigskip
{\em Physics Department,}\\
{\em  Comisi\'on Nacional de Energ\'{\i}a At\'omica,} \\
{\em Av.Libertador 8250, (1429) Buenos Aires, Argentina}

\end{center}
\vspace{.5cm}
\begin{abstract}
\noindent
We present an effective field theory formulation for a class of 
condensed matter systems with crystalline structures for which some 
of the discrete symmetries of the underlying crystal survive the 
long distance limit, up to mesoscopic scales, and argue that this class 
includes interesting materials, such as $Si$-doped $GaAs$. 
The surviving symmetries determine  
a limited set of possible effective interactions, that 
we analyze in detail for the case of  $Si$-doped $GaAs$
materials. These coincide 
with the ones proposed in the literature to describe the 
spin relaxation times for the $Si$-doped $Ga As$ materials,
obtained here as a consequence of the choice of 
effective fields and their symmetries.
The resulting low-energy effective theory is described in terms 
of three (six chiral) one-dimensional Luttinger liquid systems
and their corresponding intervalley transitions.
We also discuss the Mott transition within the context 
of the effective theory.
\end{abstract}

\vskip 1.cm

             74.20.Rp, 71.10.Pm

\vfill
\hfill April 2008
\end{titlepage}
\pagenumbering{arabic}
%

\section{Introduction}

Both electronic and spin dynamics are major research areas in Condensed Matter
Physics, specially for materials which induce strong correlating among 
electrons. In such systems, the interplay
between several effective interactions and physical degrees of freedom 
is believed to lead to interesting effects, such as the
colossal magnetoresistance and the high $T_C$ superconductivity.
The microscopic structure of these materials is complex: 
generically, these are compounds grew on
some substrate with random impurities, yielding a crystalline
structure with some amount of disorder. Moreover, the low density
of the free carriers renders often their kinetic energy comparable to the Coulomb
energy, pushing the system close to the Hubbard-Mott transition
({\it i.e.}, metal-insulator transition driven by correlations). In this regime,
the approximations leading to the usual band theory results become invalid, and
it is assumed that the electronic system is not a Fermi liquid.

The Hubbard-Mott transition has been the subject of much attention in Condensed Matter
Physics, often studied in reference to the Hubbard Model. Actually, some authors recognize 
two types of Mott transitions \cite{Giamarchi-1}: the so-called $U$-Mott transition, in which 
at a given filling, a variation of the interaction leads to a metal-insulator transition, 
and the $\delta$-Mott transition, for which the interaction is constant and 
the doping variable, leading to a commensurate to a incommensurate transition.
The physics of the transition still remains not well-understood in $d=2$ and $d=3$ spatial dimensions. 
Moreover, most of the studies used mean-field techniques, in particular, the Dynamical 
Mean Field Theory \cite{Georges-Rev}. Some more detailed information of the transition, such as its 
universality class or critical exponents, still remain to be fully established.
Recently, it  has been claimed that the Mott transition is of the gas-liquid type, 
and belongs to the Ising Universality class.\cite{Georges-Mott-Universality} \cite{Kotliar-Rozenberg-orderP}.
One may argue, however, that at least for some systems of interest, the separation between $U$-Mott and $\delta$-Mott
transitions is somewhat artificial, since, at low enough density values, changing the number of free particles 
({\it i.e.}, doping) changes simultaneously the interaction and the commensurate-incommensurate character  
of the $\delta$\ Mott transition.

On the other hand, the Mott transition has recently attracted interest in 
a rather different arena, namely, the 
non-equilibrium  magnetic properties of spin polarized electrons. In a remarkable experiment, Kikkawa and 
Awschalom\cite{Kikkawa1} have found that the spin lifetime in Si-doped $GaAs$ semiconductors can be enlarged by up
to three orders of magnitude by the introduction of $n$-doping of the order of $n=10^{-3} cm^{-3}$.
Among the early mechanisms that have been used to study the spin decay in 
semiconductors are the ones proposed by Elliot-Yafet (E-Y)\cite{Elliot} \cite{Yafet}, 
D'yakonov-Perel (D-P)\cite{D-P} , Bir-Aronov-Pikus  
\cite{B.A.P}, the relaxation due to nuclear spins, the Dyaloshinskii-
Moriya (D-M) interaction\cite{Dzyalo}\cite{Moriya} and others. Recently, 
some calculations have been repeated for the E-Y mechanism, in the low 
temperature regime \cite{Tambo}, and for the D-P mechanism \cite{Flatte} but 
they disagree with the experimental results by nearly three orders of magnitude at 
doping near to $n=10^{16}\ cm^{-1}$.
Moreover, it has been claimed that the enhancement of spin lifetime
can be understood within the D-M picture \cite{kav1}. This claim has been 
criticized by Gorkov, who has pointed out that the enhanced lifetimes take
place at doping levels near to the metal-insulator (Mott) transition
\cite{Gorkov}. In fact, while all the preceding mechanisms and calculations are 
perturbative, the Mott transition is a strongly correlated effect,thus at first
 glance, the Hiller-London calculation in ref \cite{kav1} seems misleading as 
was pointed out in ref \cite{Gorkov}.
At present, however, we are not aware of any calculations of the Mott transition for systems
like $GaAs$. In particular, it is unknown how the crystalline symmetries
and the band structure could affect the Mott transition in semiconductor
systems. The answer to these questions cannot be given by performing 
band structure calculations, due to the limitations imposed by the strong correlations.

To summarize: on the one hand, there is not as yet a consensus about the 
nature of the spin relaxation mechanism in $n$-doped semiconductors: 
several different models are all based in band structure and dilute gas scattering approximations,
which may not be valid for a phenomenon involving strong correlations. 
On the other hand, several characteristics of the Mott transition are still
not well understood, and its eventual relevance to the spin relaxation dynamics 
it is still an open issue .
A first step towards the understanding of both problems would be the formulation of a unified 
framework for describing systems of electrons including both the band structure information
and the strong correlations effects.  
The main purpose of this paper is to propose such a framework, given by an effective field theory  
that can be applied for crystalline condensed matter systems of the type discussed above.
Effective field theories (EFTs) are a general, systematic and powerful framework
to describe the low energy (long distance) dynamics of systems with
one or more energy scales. These include and unify some earlier approaches, 
such as the Landau-Ginzburg
theories and the Fermi liquid theories, among others\cite{Polchinski}. They
provide a {\it systematic tool} to analyze the different interactions
and their relative importance through the use of the Renormalization Group (RG) ideas 
and techniques. Thus, this framework is well suited for the crystalline systems in the regime
discussed above, where a naive calculation shows an  energy scale of about 
$E\sim {\hbar }/{100ps}\sim 6.5\ 10 ^{-3}\ mev$
at doping levels of $n\sim 10^{16}\ cm^{-3}$, while the typical Fermi
energy is of the order of $E_F \sim 2\ mev$ at this doping, and the gap energy
is of the order of some ev, $E_g\sim 3\ ev$.
As a test bench for the EFT, we discuss the Mott transition for samples of $Si$-doped $GaAs$, 
and show that it can be used as a useful tool for discussing some properties of the spin 
relaxation dynamics processes, although we shall not address those processes directly, given 
that 
these are non-equilibrium process and therefore, outside the applicability range of any
field theory. 

This paper is organized as follows: in the section 2 we write down the general EFT 
for electrons in crystalline systems. 
In section 3 we consider the inclusion of low disorder in the EFTs
using standard techniques. 
Section 4 is devoted to the formulation and discussion of the EFT for the 
case of $Si$-doped $GaAs$ materials, including some perturbative 
Renormalization Group calculations. 
In Section 5, we discuss the strong coupling regime of the EFT and its
relation to the Mott transition, the scenario for the
appearance of a $d$-wave density order parameter and their relevance to
the systems considered here.
Finally, we give some Conclusions.

\section{Effective Field Theories for Lattice Systems}

Condensed matter systems are quite often hard to solve, due to
their large number of degrees of freedom and to the strong
interactions that arise among them. However, the description of
many systems at a given energy scale becomes simpler in terms of a
few representative degrees of freedom. This observation, which is
a consequence of the Renormalization Group, is the key idea behind
the formulation of a given system in terms of Effective Field
Theories (EFTs) \cite{Polchinski}\cite{Shankar} . The general
scheme for applying the method of EFTs to a given system starts by
conveniently choosing the characteristic degrees of freedom that
describe the system at a given scale. In general, this choice is
suggested by the phenomenology or the experimental data. Then, one
proceeds to write down the most general local action which is
consistent with the symmetries of the system, in terms of (second
quantized) fields (or `order parameters') that represent those
degrees of freedom. However, one only retains a few of all the
possible terms in the action, by analyzing the scaling behavior of
each, {\it i.e.}, only those terms that do not decrease in the limit of
low energies or long distances (relevant and marginal terms). The
so-obtained action is then called the (Wilsonian) Effective Active,
and allows for a complete description of the system in the limit
of low excitation energies, {\it i.e.}, it yields its ground state and
low-lying excitations.

The systems we will consider here are non-relativistic electrons
moving in crystalline backgrounds. The basic field is taken to be
non-relativistic fermionic (spin $1/2$) field. The basic
symmetries of the system of electrons moving in a given background
are: i) the electron number, ii) the discrete lattice symmetries,
iii) the spin $SU(2)$ and iv) the $U(1)$ gauge symmetry of
electric charge. We would like to anticipate here that we will
consider relativistic corrections to the standard Fermi liquid
theory by allowing for spin-orbit couplings. Anticipating some of
the discussion, we would like to make the following remark: {\it
in some crystalline systems, the full $SU(2)$ symmetry group could
be reduced to subgroups of it compatible with the symmetries of
the point group of the crystal.}

The low-energy properties of the above electronic systems are
described in terms of the dynamics of the Fermi surface, which is
written in terms of the basic fermionic fields in the EFT
formulation.The ground state of the non-interacting theory consist
of a filled up Fermi sea with a Fermi surface. The elementary
low-lying excitations of the system are particle-hole fluctuations
above the Fermi surface. The geometrical shape and properties of
the Fermi surface are important characteristics which serve for
characterizing the system in a universal manner. For spherical (or
convex) Fermi surfaces, the EFT in three spatial dimensions
yields the familiar Landau theory of Fermi liquids
\cite{Polchinski},\cite{Shankar}. However, there are other cases
in which the interaction of the electrons and the crystalline
structure can modify this simple picture, and it becomes necessary
to take into account the electronic band structure.

In order to illustrate some of the basic ideas of the method for
constructing EFTs, let us first consider the following action:
\beq 
S_{0}=\int dtd^{3}x\ \psi ^{\dagger}(x,t) \left[\ i\partial
_{t}\ -\ H\ \right]\psi (x,t)\ , 
\eeq 
where $\psi (x,t)$ is a
fermionic field and $H$ the Hamiltonian of the system, which plays
an important role in our non-relativistic systems. The field can
be expanded in a Fock space basis as follows:
\beq 
\psi (x,t)\ =\ \sum _{n}\ \phi _{n}(x,t)\ c_{n}\ , 
\eeq
where $c_{n}$ ($c_{n}^{\dagger }$) are the fermionic annihilation
( creation) Fock space operators, respectively, and $\phi_{n}(x,t)$
are the corresponding wave functions. The Euler-Lagrange equations
of motion are then Schrodinger equations: \beq \left[\
i\partial_{t}\ -\ H\ \right]\phi _{n}(x,t)\ =\ 0 \label{eqsch}\eeq

For time-independent Hamiltonians and translation-invariant
crystalline backgrounds, the solutions to (\ref{eqsch}) may be
chosen in the Bloch form:
\beq 
\phi_{n}(x)\ =\ {\rm e}^{i{\bf k}\dot {\bf r}}\ u^{(n)}_{{\bf
k}}({\bf r})\ , 
\eeq 
where ${\bf k}$ is the wave-vector, which
belongs to the first Brillouin zone. The spatial periodicity then
gives rise to the band structure, defined by:
\beq 
H\phi _{n}({\bf k})\ =\ \epsilon_{n}\phi _{n}({\bf k})\ .
\eeq
A functional integral of the fields that describes the system is
given by the partition function:
\beq Z\ =\ \int D\psi^{\dagger }D\psi\ \exp\left(i\sum _{n}\int dt\int
_{\Gamma_1}d^{3}k\ \psi _{n}^{\dagger }({\bf k},t)\left[i\partial
_{t}-\epsilon _{n}({\bf k})-\mu \right]\psi _{n}({\bf k},t)\right)\ ,
\label{zfull} \eeq where $\mu $ is the chemical potential, and the
$\Gamma_1$ denotes the first Brillouin zone.

We would now like to construct the EFT corresponding to
(\ref{zfull}). We consider a characteristic energy scale
$E_{\Lambda }$, and a corresponding set of wave momenta 
at the Fermi surface $ {\bf k}$ such that $|{\bf k}| <{\Lambda }$. 
Here $\Lambda $ is a chosen cut-off scale that serves
to divide the fields into their high and low frequency components:
$\psi =\psi _{L}+\psi _{H}$. More precisely, $\psi _{L} ( {\bf
k},t)$ depends only of those momenta ${\bf k}$ satisfying $|{\bf
k}| \leq |{\Lambda }|$. The low-energy EFT is obtained by formally
integrating out the high frequency components in (\ref{zfull}):
\beq Z\ =\ \int D\psi _{L}^{\dagger }D\psi _{L}\ {\rm
e}^{iS_{\Lambda }(\psi _{L}\psi _{L}^{\dagger })}\ , \label{zeta1}
\eeq where \beq {\rm e}^{iS_{\Lambda }(\psi_{L},\psi_{L}^{\dagger
})}\ =\ \int D\psi_{H}^{\dagger }D\psi_{H}\ \exp(iS_{0}
(\psi_{L}^{\dagger },\psi_{L},\psi_{H}^{\dagger },\psi_{H}))\ .
\eeq
$S_{\Lambda }$ is known as the low energy or Wilsonian effective
action. It can be expanded in the space of all theories in the
(infinite) basis of all local operators $ O_{n}$ allowed by the
symmetries: 
\beq S_{\Lambda }=\int dt\int d^{3}k\ \sum _{n}\ O_{n}\ . 
\eeq
Notice that this effective action may contain free and (local)
interacting terms among the chosen effective fields. For the case
at hand, given by (\ref{zeta1}) we obtain:
\beq Z\ =\ \int D\psi_{L}^{\dagger }D\psi _{L} \exp\left( i\sum _{n}\int
dt\int_{\Gamma_{\Lambda}} d^{3}k\ \psi _{_{L}n}^{\dagger }({\bf
k},t) \left[i\partial_{t}\ -\ \epsilon _{n}({\bf k}_{L})-\mu
\right]\psi _{_{L}n}({\bf k},t)\right)\ , \eeq
where $\Gamma_{\Lambda}$ is not the entire Brillouin zone, but
only the region near its center satisfying $|{\bf k}| \leq
|{\Lambda }|$. Therefore, all momenta appearing from now on are
small, and it is possible to expand any function of momenta, like
the energy, in a power series as follows:
\beq \epsilon _{n}({\bf k})\ =\ \alpha _{i}\ k_{i}\ +\ \alpha
_{ij}\ k_{i}k_{j}\ +\ \beta _{ijk}\ k_{i}k_{j}k_{k}\ +\
\tilde{\beta }_{ijk}\ k_{i}k_{j}k_{k}\sigma _{i}\ +\ \dots\ ,
\label{tensores} \eeq
where the Greek quantities denote coefficients and the Latin
indices ran among the three spatial directions ($i=1,2,3$).
The different coefficients appearing in
(\ref{tensores}) are actually spatial tensor structures, which can
be classified by the spatial symmetries\footnote{Note that only
third order tensors can couple the spin operator, because these
may transform as a pseudo-vector in one index and as a vector in
the remaining two, giving rise to a scalar when contracted with
the spin pseudo-vector} of the system. In fact, it is well known
that near the center of the Brillouin zone, the electronic wave
functions transform as the irreducible representations of the
point group of the crystal.

In the following, we will focus on the EFT only, and therefore
will suppress the subindices $\Lambda$ and $L$ in the effective
action and effective fields, respectively. Therefore, for low
energy and large distances the free part of the effective action
may, at the $n-th$ band, be written as: \beq S_{0}^{(n)}\ =\ \int
dt\int d^{3}k\ \psi _{n}^{\dagger }({\bf k},t) \left[i\partial
_{t}-\sum _{\alpha }\hat{E}^{(n)}_{\Gamma _{\alpha }} (k_i
,k_{j},k_{k}......)\ -\ \mu\right]\psi _{n}({\bf k},t)\ , \eeq
where $_{\Gamma _{\alpha }}$ runs over all the different
representations of the point group and the energy operators
$\hat{E}$ have to be expressed in terms of the basis functions
within each irreducible representation. This generalizes the case
considered above for more complex Fermi surfaces, that may display
less familiar dispersion relations and several components. The
elementary excitations are particle-hole pairs above the Fermi
surface, which is defined by the equation $\epsilon_{n}({\bf
k})=\epsilon_F$, whereas $\epsilon_F$ is the Fermi energy. 
For spherical Fermi surfaces, these elementary excitations are
Landau quasiparticles (dressed electrons) as has been pointed out in
\cite{Shankar}. The fact that the Landau theory of Fermi
liquids is an EFT has been discussed in detail in \cite{Polchinski}.  
In order to separate the different scaling components of the momenta,
we consider the decomposition ${\bf k} = {\bf k}_0+{\bf p}$,
whereas ${\bf k}_0$ is a vector lying on the Fermi surface and
${\bf p}$ is a vector orthogonal to it. The scaling down of the
energy $E\to sE$, with $s \to 0$ a small dimensionless parameter,
induces a corresponding scaling on the different components of the
momenta such that $|{\bf k}|_{0}\to |{\bf k}|_{0}$ and $|{\bf
p}|\to s |{\bf p}|$. The energy operators are constructed
accordingly:
\beq \tilde{E}_{\Gamma _{\alpha }}(k_i,k_{j},k_{k},\dots )\ =\ \hat{E}_{\Gamma _{\alpha
}}(k_{0i},k_{0j},k_{0k},\dots)\ +\ \frac{\partial \hat{E}_{\Gamma
_{\alpha }(k_{0i},k_{0j},k_{0k},\dots)}}{\partial k_{n}}\ p_{n}
\label{operadores} \eeq
There is a set of different Fermi
velocities given by the last term of the R.H.S of
(\ref{operadores}). 

We now turn to the interactions.
Note first that the aim of the EFT approach is not to {\it deduce} the 
form of the interaction terms in the effective theory (
{\it i.e.}, the interactions among the effective degrees of 
freedom)  by direct transformation of an underlying 
microscopic interaction, but rather to make a judicious 
{\it ansatz} of the possible interactions in terms of the
effective field symmetries (for a detailed discussion
see \cite{Polchinski}). 
At zero temperature, $\mu=\epsilon_F$ and the
dynamics of the low-lying excitations ({\it i.e.}, the
EFT fields defined as discussed above) above the ground state is
given by:
\beq S_{0}\ =\ \int dt\int d^{3}k\ \psi _{n}^{\dagger
}({\bf k},t)\left[i\partial _{t}\ -\ \sum _{\alpha
}\lambda_{\alpha}\frac{\partial{E_{\Gamma _\alpha
}(k_{0i},k_{0j},k_{0k},\dots)}}{\partial{k_i}}\ p_i\right]\psi
_{n}({\bf k},t)\ , \label{sfree} \eeq
where the values of the dimensionless parameters
$\lambda_{\alpha}$ are not predicted by the EFT, but are expected
to be of order $1$ (`naturality' \cite{Polchinski}). 
If one tries to go beyond the scope of the EFT
formulation, and insist in matching it with an underlying
short-distances theory, then one may conclude that they depend on
the details of the band structure and may be exactly zero only if
there is an additional symmetry, or very small if the band is very
far from the energy scale one is considering.

The interaction terms may also be included in the EFT in a similar
fashion. For convenience, we consider here the fields as functions
of frequency, rather than the time coordinate (using the obvious
notation $ (1)=(\omega_1,\bf{k}_1)$):
\beq
 S_{int}= \sum _{n_{i}}\int\ d\mu
 \left[\psi _{n_{1}}^{\dagger }(1)\psi _{n_{2}}^{\dagger }(2)\psi _{n_{3}}
\psi _{n_{4}}(4)\ V({\bf k}_{1},{\bf k}_{2},{\bf k}_{3},{\bf
k}_{4})\ \bar{\delta }_k \delta _\omega \right] \label{sint} \eeq
with the integration measure $d\mu=d\omega_{1}d^{3}k_{1}d\omega_{2}
d^{3}k_{2}d\omega_{3}d^{3}k_{2}d\omega_{4}d^{4}k_{4}$ and  the
short hand notation: $\bar{\delta }_k \delta _\omega= 2\pi
\bar{\delta }(k_{1}+k_{2}-k_{3}-k_{4})\delta (\omega _{1}+\omega
_{2}-\omega _{3}-\omega _{4})$ where $\bar{\delta}$ is a
delta-function enforcing momentum conservation mod $2\pi/a$ ($a$
is the cubic lattice spacing), which allows for Umklapp process as
well: 
\beq {\bf k}_{1}\ +\ {\bf k}_{2}\ -\ {\bf k}_{3}\ -\ {\bf
k}_{4}\ +\ {\bf Q}\ =\ 0\ , \eeq 
where ${\bf Q}=m(\pi/a,\pi/a,\pi/a)$ and m is an integer .
Including higher interaction terms is not necessary due to 
the more and more irrelevant character of them \cite{Polchinski}.

The complete EFT action is thus obtained from (\ref{sfree}) and (\ref{sint}),
and it is formally given by the following partition function:
\begin{eqnarray}
Z&=&\int D\psi ^{\dagger }D\psi \exp\left\{i \int dtd^3k 
\psi _{n }^{\dagger }({\bf k} ,t)\left[i\partial _{t}- 
\sum _{\alpha }\lambda_{\alpha}\frac{\partial{E_{\Gamma _\alpha }(k_0i)}}
{\partial{k_i}}p_i\right]\psi _{n,}({\bf k},t)\ + \right . \nonumber\\
&& \left . \sum _{n_{i}} \int d\mu \left[\psi _{n_{1}}^{\dagger }(1)
\psi _{n_{2} }^{\dagger }(2)\psi _{n_{3} }(3)\psi _{n_{4} }(4)
V({\bf k_1},{\bf, k_2},{\bf k_3},{\bf k_4})\right]\bar{\delta }_k' 
\delta _\omega  \right\} \nonumber\\
\end{eqnarray}
At this point, we emphasize that the field operators 
$\psi _{n,}({\bf k},t)$ do not represent the original
(free) electrons, but rather the effective (interacting) ones. 

\section{Effective Field Theory for Impurity Systems}

In this section we would like to discuss the effects of impurities
and doping in the systems we have considered in the previous
section. The approach we take here is a standard one, based 
on the quantum mechanical established procedures which
are promoted onto the field themselves by considering them
as a linear superposition of the second-quantized 
operators acting on a Fock space, with
amplitudes given by the modified wave functions.
Quite often, impurity effects lead to interesting phenomena,
such as in the case of high $T_c$ superconductors in underdoped
cuprates. 

The inclusion of impurities makes necessary to reformulate the EFT, since they 
break translation invariance, so that Bloch wave functions are not solution 
of the Schrodinger equation of the crystal. However, since in the EFT formulation 
we may to choose the 
relevant degrees of freedom, we consider the 'impurity orbitals' as the new basis for the wave 
functions appearing in the field definitions. 
They have the important property that can be extended to the {\it mesoscopic domain}.
In fact, as is well known, for a single impurity the Wannier
theory of shallow donors shows that the hydrogen -like orbitals of
typical impurities in semiconductors have an effective Bohr radius
of about 50 \AA  up to 100 \AA . In order to show how this
mesoscopic scale arises and how it matches the EFT framework, we
would like to recall briefly the Wannier theory of shallow impurities
\cite{Ziman} \cite{Cardona} \cite{Ashcroft}. Let us consider an
impurity characterized by the  potential $\delta V $, within a perfect
crystal (quantities refer to it with a $0$ subfix).  A solution of the time
dependent Schrodinger equation
\beq
\left[H_0+\delta V({\bf r})\right]\phi({\bf r}, t)=i\hbar\partial _t \phi({\bf r},t)
\eeq
may be expanded in the Wannier representation
\beq
\phi({\bf r},t)=\sum_{n,l} f_n({\bf r_l},t)a_n({\bf r}-\bf{r_l})
\eeq
where $ a_n({\bf r}-\bf{r_l})$ are Wannier functions ({\it i.e.},
discrete Fourier transforms of the Bloch functions) which are
solutions of the pure crystal electronic problem, $\bf{r_l}$ is
the real space position of the atom, and $f_n(r_l,t)$
play the role of enveloping functions. After a standard
transformation, this leads to the equation (for details see
\cite{Cardona}):
\beq
\left[ \epsilon_n(-i\nabla_l)+\delta V(r)_{nl,n'l'}\right] f_n({\bf r_l},t)\ =\ 
i\hbar\ \partial _t f_n({\bf r_l},t)\ ,
\eeq
where $ \epsilon_n(k)$ are the band energies of the perfect crystal, $\nabla_l$ 
is the discrete gradient with respect to the 
impurity site coordinates, and $\delta V(r)_{nl,n'l'}$ is the matrix element of 
the impurity potential between different Wannier functions. Now $ \delta V(r)$ 
may be expanded around $r_l$ , under the assumption of a slowly variating potential, 
$a_0|\nabla \delta V(r)|\ll \delta V(r)$ and this equation becomes
\beq
\left[\epsilon_n(-i\nabla_l)+\delta V(r_l)\right] f_n({\bf r_l},t)\ =\ 
i\hbar\ \partial _t f_n({\bf r_l},t) \label{envolvente}
\eeq
Note that equation (\ref{envolvente}) does not involve the
Hamiltonian of the perfect crystal $H_0$. Knowing $\epsilon({\bf
k})$ we can solve for $f_n(\bf {r},t)$ in a continuous
approximation. Since  $\epsilon({\bf k})$ contains the symmetries
of the lattice, this will also occur with $f_n$, provided $\delta
V$ is small. If, as it occurs in a semiconductor, the band energy is still
parabolic and the potential corresponding to a single 'shallow
impurity' then the  equation becomes the Schrodinger equation of
an Hydrogen system, except that the Coulomb potential is replaced
by an screened effective Coulomb potential, whose
dielectric constant is modified by the presence of the host
crystal $ V_s=e/{4\pi\epsilon r}$. The change
in the dielectric constant gives raise to new spacial scale for the
wave functions, and the effective Bohr radius becomes of the order of 100 \AA,
well within the mesoscopic scale.

To write down a field theory, let us consider an 'arrange of single impurities'.
The degrees of freedom we chose are associated to the impurity level, so that 
the enveloping wave functions will be a suitable basis,{\it i.e.}, the field operators 
are:
\beq
\psi({\bf r},t)=\sum_n c_n f_n({\bf r},t)
\eeq
where $c_n$ and $c^{\dagger}_n$ are the annihilation and creation
operators of electrons in the effective bands whose wave functions
are $f_n$. An equation like (\ref{envolvente}) may be considered as
the Euler-Lagrange equation corresponding to the following action:
\beq
S_0=\int dtd^3k\ \psi _{n}^{\dagger }({\bf k},t)\left[i\partial_{t}-\epsilon_{n}
({\bf k_l})-\delta V({\bf x})- \mu \right] \psi _{n}({\bf k},t)
\eeq
where  $\delta V({\bf x})=\delta V({\bf x}+{\bf R})$ plays the role
of a `lattice potential', now defined over the {\it mesoscopic domain}, which
is invariant under the discrete translation symmetry. Since the
functions $\epsilon_n({\bf k})$ are the exact energies of the pure
crystal, they contain the microscopic symmetries that near ${\bf
k}\sim 0$ are those of the point group, In this way a single
impurity extend these symmetries over mesoscopic domain. In this
way we have recovered the discrete symmetry, so the electrons can
exchange discrete momentum with the `impurity lattice'. Note that
$\delta V $ transforms as the temporal component of a gauge field
$A_0$:
\beq
S_0=\sum_n\int_{\Omega_1} dtd^3k \psi _{n}^{\dagger }({\bf k},t)
\left[iD_{0}-\epsilon_{n}({\bf k_l})- \mu \right] \psi _{n}({\bf k},t)
\eeq
where $D_0=\partial_t-iA_0$\ .
In real samples, however, impurities are not localized in a perfect 
spatial arrange,
but are randomly distributed in a complex way that depends
on the impurities, on the host crystal and on the growing process, so
that the behavior of the EFT could depend on some of this data. Here we shall
consider only the case of {\it substitutional shallow impurities}, such as
those that appear in samples grown by molecular beam epitaxy
(MBE). For this case, we have a `low degree of disorder'; since
the impurities are placed  in some of the  sites of the
crystal lattice, the Hamiltonian of the problem can be modeled by:
\beq
H_g=H_0\ +\ g({\mathbf \zeta_i})\ \delta V({\bf x_i})\label{Hamilt-desord}
\eeq
with $g({\mathbf \zeta _{i}})=1$ if ${\bf x_i=\mathbf \zeta
_{i}}$ and $g({\bf \zeta _{i}})=0$ if ${\bf x_i\neq \zeta _{i}}$,
and $\zeta_i$ denote the impurity position.
Now we have to solve the eigenvalue problem:
\beq
H_{g}|\phi _{g}\rangle =\epsilon _{g}|\phi _{g}\rangle
\eeq
where $g$ acts as a parameter that takes values in a stochastic fashion with some
definite probability distribution $P(g)$. It follows that the set ${H_g}$ may be
considered as a ensemble of Hamiltonians in the random matrix theory (RMT) sense,
with a probability distribution $\mu (H)$ induced by $P(g)$.
As is well know in RMT, the statistical properties of the ensemble are fixed by the
symmetry properties of the Hamiltonian ensemble. Moreover, the classical Wigner ensembles are
defined by invariance requirements \cite{RMT-Rev-Alemanes}: given a particular member $H$ in the ensemble,
all the others are generated by similarity transformations on the Hilbert spaces. This postulate
ensures that there is no preferred basis on the Hilbert space.
Following this line of thought, we assume that the symmetry of the random matrix ensemble 
is the same as that of the EFT.
Physically, this means that we shall assume that the {\it impurities form an array}, such  
that the symmetries of the crystalline structure are extended onto the mesoscopic domain. 
A similar mechanism has been argued in \cite{Lauglin-Nayak} for the case of cuprate 
superconductors, in the sense that the $d$-wave order parameter survives to the presence 
of disorder. 
We shall discuss briefly the applicability of this hypothesis to the case of 
$Si$-doped $GaAs$ in the next section.

A more precise formulation of the previous ideas starts by considering equation 
(\ref{Hamilt-desord}) for fixed $g$, which defines the action:
\beq
S_{0g}=\int dtd^3k \psi _{n}^{\dagger }({\bf k},t)\left[i\partial_{t}-
gA_0-\epsilon_{ng}({\bf k})-\mu \right] \psi _{n}({\bf k},t)\ .
\eeq
Following the same steps that yielded (\ref{tensores}), we obtain:
\beq
\epsilon _{n,g}(k)=\alpha^g _{i}k_{i}+\alpha^g
_{ij}k_{i}k_{j}+\beta^g _{ijk}k_{i}k_{j}k_{k}+\tilde{\beta^g
}_{ijk}k_{i}k_{j}k_{k}\sigma _{i}\label{tensores-2} 
\eeq
Since we have assumed that within each array the lattice symmetries 
survived the change of scale, the energy
operators are still defined by the irreducible representations of the point group of the
 crystal, so that the effect of the disorder appears throughout the coupling constants 
only. The partition function $Z_{\lambda}$ for the theory parametrized by a particular
value of $\lambda$ is given by:
\begin{eqnarray}
Z_{\lambda }=\int D\psi D\psi ^{\dagger }e^{i \int dtd^{3}k 
\psi _{}^{\dagger }({\bf k},t)\left[i\partial _{t}+\lambda_0A_0 -
\sum \lambda _{\alpha }\hat{E}_{\Gamma _{\alpha }}(ki,k_{j})\right]\psi({\bf k},t)}
\end{eqnarray}
To complete the definition of the EFT, we must average over the disorder
parameter $\lambda$. This can be done in at least two cases:
i) quenched disorder, which for static or thermodynamics theories 
is often handled thorough the replica trick. Such is the case,{\it e.g.}, of the Sherrington-Kirpatrick model
of spin glass.or the field theoretical calculations of the average density of states
in the Anderson localization problem.\cite{RMT-Rev-Alemanes} \cite{Lee-Ramakri} .
ii) dynamical theories of sp\'{\i}n glasses avoid the use of the replica trick.
In these theories, the generating functional is normalized \cite{Sompolinsky} \cite{Cugli-Lozano-1} 
and the quenched average can be done directly using the generating  functional. We will present 
here an approach to averaging over the dynamical theory that is inspired in the 
'static formalism' and follows closely the approach presented in \cite{Zee}.

Let us consider the following generating functional, for a fixed
set  $\{{\lambda _0,\lambda_{\Gamma _{\alpha}}}\}$:
\beq Z_{\lambda }=\int D\psi D\psi ^{\dagger }\exp( i \int
dtd^{3}k\left\{ L_{eff}+\psi _{}^{\dagger }({ \bf k},t)J(t)+\psi
_{}({\bf k},t)J(t)^{\dagger }\right\}), \eeq
with
\beq L_{eff}= \psi _{}^{\dagger }({\bf k},t)\left[i\partial
_{t}+\lambda_0A_0-\sum \lambda _{\alpha }\hat{E}_{\Gamma _{\alpha
}}(ki,k_{j},..)\right]\psi ({\bf k},t) \eeq
For fixed $\lambda $, the Green's functions are defined by:
\beq
G_{\lambda }(t-t',{\bf k})=\langle T(\psi ^{\dagger }({\bf
k},t)\psi ({\bf k},t')\rangle\ ,
\eeq
and the average over the disorder (denoted by double brackets) can be done using:
\beq \langle \langle G_{\lambda }(t-t')\rangle \rangle  =\langle
\langle \frac{\delta \, \ln Z_{\lambda }(0,0)}{\delta J(t')\delta
J(t)}\rangle \rangle \eeq
where $G_{\lambda }(t-t')=\int d^{3}kG_{\lambda }(t-t',k)$ . 
We now apply the replica trick,{i.e.}, the identity
\beq \
\ln Z=lim_{n\rightarrow 0}\left(\frac{Z^{n}-1}{n}\right) \ .
\eeq
The average Green's function ( denoted
by  $G_{\bf k}(\tau)$ with $\tau=t-t'$) becomes
\begin{eqnarray}
G_{\bf k}(\tau)&=& \partial _{Jt}\partial
_{Jt'}lim_{n\rightarrow 0}\frac{1}{n}\int \prod _{a=1}^{n}{}D\psi
_{a}D\psi _{a}^{\dagger }d\lambda P(\lambda )e^{i S[\lambda]} \\
 S[\lambda]&=&{\int dtd^{3}k\psi
_{a}^{\dagger }\left[i\partial _{t}-\lambda _0 A_0-\sum \lambda
_{\alpha }^{a}\hat{E}_{\Gamma _{\alpha }}\right]\psi _{a}+J\psi
_{a}^{\dagger }+J^{\dagger }\psi _{a}}
\end{eqnarray}
where we have used $ \partial _{Jt}\partial _{Jt'}=\delta^2 /\delta J(t')\delta J(t)$. 
We consider a standard Gaussian probability distribution for the coupling constants,
\beq
P(\lambda )\ =\ \frac{1}{\sqrt{2\pi }\sigma }\ e^{-(\lambda
-\lambda _{0})^2/(2\sigma ^{2})}
\eeq
so that the integration over disorder is Gaussian, and can be done 
in a closed way:
\begin{eqnarray}
&&G=  \partial _{Jt}\partial _{Jt'}lim_{n\rightarrow 0}\frac{1}{n}\hspace{10 cm}\nonumber\\
&&\left\{ \int \prod _{a=1}^{n}D\psi _{a}D\psi _{a}^{\dagger } \exp{\int dtd^{3}k
\psi _{a}^{\dagger }\left[i\partial _{t}-\lambda_0 A_0-
\sum \lambda _{0\alpha }^{a}\hat{E}_{\Gamma _{\alpha }}\right]\psi _{a}
+J\psi _{a}^{\dagger }+J^{\dagger }\psi _{a}} \right. \nonumber \\
&& \exp{ \frac{\sigma ^{2}}{2}\sum _{\alpha \beta }\int dt_{1}dt_{2}d^{3}k_{1}d^{3}k_{2}
\psi _{a}^{\dagger }(1)\psi _{a}(1)E_{\Gamma \alpha }^{a}(1)\, 
\psi _{b}^{\dagger }(2)\psi _{b}(2)E_{\Gamma _{\beta }}^{a}(2))}\nonumber\\
&&\left.\exp \sum _{n_{i}} \int d\mu\psi _{n_{1}}^{\dagger
}(1)\psi _{n_{2} }^{\dagger }(2)\psi _{n_{3}}(3)\psi _{n_{4}
}(4)V({\bf k_1},{\bf k_2},{\bf k_3},{\bf k_4})\bar{\delta}_k \delta_{\omega}\right\} \label{Gmean}
\end{eqnarray}
where we have included the interaction term,
since we assume that the Coulomb potential is independent of the disorder. 
In the second line of (\ref{Gmean}), we have used the short-hand 
notation $E_{\Gamma_0}=A_0$. Note that the impurity potential can mix 
with other operators.

Thus, integrating out the random disorder we have introduced an effective interaction 
between operators with different symmetries. Effective interactions arising from 
disorder were already present in early models, such Sherrington-Kirkpatrick model, and 
in the study of the classical dynamic of spin glasses \cite{Sompolinsky}. In the
problem at hand, they acquire a new meaning. 
In fact, it has been argued that in disordered or complex condensate matter systems, 
different symmetries and order parameters may compete\cite{Dagotto}. In the problem
we are considering, this behavior results directly from EFT: we have began with a set of
definite degrees of freedom and given symmetries but the process of averaging over the
disorder yields a theory mixes those degrees of freedom and new induced interactions,
which compete with the Coulomb interaction, arise among them. For the rest of the paper, 
we will consider the case of low disorder only,{\it i.e.}, $\sigma $ small so that 
the effective induced interactions are negligible in comparison to the Coulomb term.

\section{Effective Field Theories for systems of $Si$-doped $GaAs$ compound materials}

In this section, we  apply the framework of the preceding
one to analyze the behavior of the $Si$-doped $GaAs$ samples 
employed in the spin relaxation experiments of
reference \cite{Kikkawa1}. We obtain the 
effective Hamiltonian and analyze the RG lowest order 
perturbative corrections to the coupling constants values. 

Let us first consider an array of $Si$ impurities in $GaAs$ with
the symmetries of the substrate. This hypothesis is reasonable,
because $GaAs$ has zincblende crystalline structure and the silicon
grown in fits in a diamond lattice (with the same structure of the
zincblende, but with only one kind of atom in the basis). The
samples of $Si$-doped $GaAs$ considered are grown by Molecular 
Beam Epitaxy, which implies that the impurities are almost all 
substitutional. Moreover, the
single-impurity wave functions have sizes of about 100 \AA , and for
doping near $ 10^{17} cm^{-3}$, these are separated by distances of 
about 200 \AA. In addition, some self-organized structures
and three dimensional nanowires have been observed inside samples
of $Si$-doped $GaAs$ \cite{Si-GaAs-Selforganized} .
Note that this hypothesis may or may not be valid for systems
different to the one considered here. A general rule as to how to 
proceed cannot be formulated at this point and the specific 
properties of each system should be taken into account (as is often the case 
in the framework of EFTs). 

From the symmetry point of view, the zincblende belongs to the
spatial group $T_{d}^{2}$ or $F_{\bar{4}3m}$  \cite{Cardona}, which has 
$24$ elements (it is the same as the point group of the 
tetrahedron, $T_{d}$ ). There are $8$ $C_{3}$ rotations about 
the diagonals $(\pm 1,\pm 1,\pm 1)$, 3 $C_{2}$ rotations about
the axis $(1,0,0 ) , (0,1,0)$ and $(0,0,1)$, and 6 $C_{4}$ improper 
rotations about $ (1,0,0 ) , (0,1,0)$ and $(0,0,1)$. The group $T_{d}^{2}$, has 
also six reflections $\sigma$ with respect to the planes 
$[1,1,0],[1,\bar{1},0],[1,0,\bar[1]],[011]$ and $[01\bar{1}]$. 
On the other hand, Silicon belongs to the spatial group $O^7_h$ 
(the diamond group), which is non-symorphic and has three glide planes 
defined by $x=a/8$, $y=a/8$, $z=a/8$ and screw axis along the directions $x,y,z$.

Ii is well known that electrons in $Si$ are located in impurity levels 
near the conduction band (donor levels) (and the holes are near the top of the valence band, 
acceptor levels) There are six Fermi points or `pockets' along the directions $(\pm 1,0,0)\pi /a^{*}$
,$(0,\pm 1,0)\pi /a^{*}$,$(0,0,\pm 1)\pi /a^{*}$, with energies:
\beq
\epsilon (k)=\epsilon _{c}+\frac{\hbar }{2}\left(\frac{k_{l}^{2}}{m_{l}}+
\frac{k_{t1}^{2}}{m_{t}}+\frac{k_{t2}^{2}}{m_{t}}\right)
\eeq
where $m_l$ and $m_t$ are the effective masses in the longitudinal and transverse directions.
As discussed in section 2, in order to write down the EFT one considers
the low-lying excitations above the ground state, which can be done (at lowest order in momentum) 
by linearizing the dispersion relation about the pocket points yielding three (six chiral) quasi one-dimensional 
gapless systems (Luttinger liquids, see Figure 1). However, as we shall soon see, there are more allowed terms 
in the EFT besides these ones.

In order to obtain all the terms, we apply the approach
of section 2 and construct the energy
operators $ \hat{E}_{\Gamma _{\alpha }} $, which transform as the
irreducible representations of the symmetry group of the system.
In our case, we consider energy operators 
invariant under the symmetries of both Silicon and $Ga-
As$,{\it i.e.}, diamond and zincblende structure.But since 
$T_d$ is a subgroup of the diamond group, all the symmetries needed to
construct the EFT are in fact contained in the group $T_d$, which has the
following characters and basis functions table:
\begin{center}
\begin{tabular}{|c|c|c|c|c|c|c|}
\hline
&
E&
3$C_2$&
6$S_{4}$&
6$\sigma $&
8$C_{3}$&
Basis Functions\\
\hline
\hline
$A_{1}$&
1&
1&
1&
1&
1&
$k_x.k_y.k_z$\\
\hline
$A_{2}$&
1&
1&
-1&
-1&
1&
$k_x ^{4}(k_y^{2}-k_z^{2})+k_y^{4}(k_z^{2}-k_x^{2})+k_z^{4}(k_x^{2}-k_y^{2})$\\
\hline
E&
2&
2&
0&
0&
-1&
\{$(k_x^{2}-k_y^{2}),2k_z^{2}-(k_x^{2}+k_y^{2})$\}\\
\hline
$T_{1}$&
3&
-1&
1&
-1&
0&
\{$k_x(k_y^{2}-k_z^{2}),k_y(k_z^{2}-k_x^{2}),k_z(k_x^{2}-k_y^{2})$\}\\
\hline
$T_{2}$&
3&
-1&
-1&
1&
0&
\{$k_x,k_y,k_z$\}\\
\hline
\end{tabular}
\end{center}
Note that only the class $T_1$ transform in a pseudo-vector
representation, so it must couple to the spin variable
to produce a scalar term in the effective action.

Using only the equations (\ref{operadores}), the character table
and neglecting four order terms ({\it i.e.}, those arising from the $A_2$
class), we obtain the effective Hamiltonian $H_{0}^{eff}= H_p\ +\ H_{so}$, where the second
term is the spin-orbit Hamiltonian:
\vspace{1 cm}
\begin{eqnarray}
H_p&=&\lambda _{0}A_0+\lambda_1\left[p_{x}+p_{y}+p_{z}\right]
+
\sum _{ç\alpha =\pm } \left[\lambda_2 (\mathbf{g}_{x\alpha }.\mathbf{p})+ \lambda_3
(\mathbf{g}_{y\alpha }.\mathbf{p})+\lambda_4 (\mathbf{g}_{z\alpha }.\mathbf{p})\right]\nonumber\\
&&\qquad \quad+\lambda_{5}\bigg[k_{ox}\,p_{y}p_{z}\,+\,k_{oy}\,p_{x}p_{z}\,
+\,k_{oz}\,p_{x}p_{y}\bigg]\nonumber\\
H_{so}&=&\lambda _{6}\bigg[k_{0x}(p_{y}^{2}-p_{z}^{2})\sigma
_{x}+k_{0y}(p_{z}^{2}-p_{x}^{2})\sigma
_{y}+k_{0z}(p_{x}^{2}-p_{y}^{2})\sigma
_{z}\bigg]\ 
,\label{EffectH}\\\nonumber
\end{eqnarray}
where ($\lambda_k$, $k=1,\dots,6$) are coupling parameters, 
$g_{x,\pm }=(0,1,\pm 1)$, $g_{y,\pm }=(1,0,\pm 1)$, $g_{z,\pm }=(1,\pm 1,0)$,
which are vectors in the reciprocal space. Note that these vectors appear also in the pseudo-potential
theory for $GaAs$ \cite{Cardona} . In (\ref{EffectH}) one may consider the 
terms linear in momenta as subsystems of quasi one-dimensional Luttinger liquids. 
There are also two parity-breaking terms: the spin-orbit part has the form
of the Dyakonov-Perel Hamiltonian (also called Dressehauss term) \cite{D-P}.
Note that term of the form $(k_{oy}^2 - k_{oz}^2 ) p_x \sigma_x $,  and the
ones obtained from it by cyclic permutations could also be considered in $ H_{so}$.
However, they do not contribute to $ H_{so}$ for the following reasoning: 
a parity transformation acting on the low-lying states of the theory not only
transforms the components of the electron momentum ${\bf p}$, but also 
exchanges two of the corresponding pocket ground states of Figure 1. 
The parity transformation reflecting the axis $x$ onto minus itself can be
seen as an exchange of the axis $y$ with $z$, while keeping $x$ fixed.
Therefore, the term $(k_{oy}^2 - k_{oz}^2 )$ changes sign, as well as
$ \sigma_x$, while $p_x$ does not. However, this term is identically
zero since $k_{oy}^2 = k_{oz}^2$. Moreover, one could also consider 
a chiral component of each Luttinger liquid ({\it i.e.}, one of the six pockets
of Figure 1), and after adoption of a convenient redefinition of the parity
transformation on this system conclude that these terms neither arise 
in this case.

\begin{figure}[ht]
\centering \leavevmode
  \epsfxsize 2.5 in
  \epsfbox{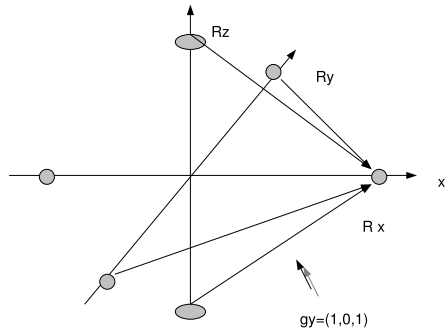}
   \caption{Ground state structure of the EFT in terms of Luttinger liquid `pockets'.}
\label{Pockets}
\end{figure}

The field theory corresponding to the Hamiltonian (\ref{EffectH}) is 
defined by its partition function, given by:
\beq
Z=\int D\psi _{L}^{\dagger }D\psi _{L}{\rm e}^{i S} \ .
\eeq
The free part of the action is given by
\beq S_{0}=\sum _{i,\alpha}\int _{-\Lambda }^{\Lambda }
\frac{dp_{\alpha}}{2\pi}\frac{d\omega}{2\pi } \left\{ \psi_i
^{*}(\omega ,p_{\alpha})\left[i\omega -\lambda_0 A_0-\lambda
p_{\alpha}+O(p^{2})\right]\psi_i (\omega, p_{\alpha})\right\} \eeq
where the index $i$ runs over the six Fermi points in the different spatial directions
({\it i.e.}, $i=R_x,L_x,R_y,L_y,L_z,R_z$, where $R, L$ mean 
left and right Fermi points,
respectively), and $\alpha $ is an index that indicates the component of the momentum along 
the six different axis defined by the directions $x,y,z,g_x,g_y,g_z$. By abuse of notation,
we take $\alpha=x,y,z,g_x,g_y,g_z$. Note that the two indices $i$ and $\alpha)$ 
are not independent; for example, if $i=R_x$ then  $\alpha$ must take on of 
the values given by $x,g_y$ or $g_z$. In Figure (\ref{Pockets}) we indicate 
each of the six possible Luttinger ground states (`pockets') and a few
possible intervalley processes produced by the effective Hamiltonian (\ref{EffectH}).

The general form of the interaction term is
\bea S_{I}=\frac{2\pi}{2!2!}\sum _{i_1,i_,i_3,i_4}\int d\mu
\psi_{i_1} ^{\dagger }(1)\psi _{i_2}^{\dagger }(2)\psi_{i_3}
(3)\psi_{i_4}(4)V_{i_1,1_2,1_3,1_4}(k_{01},k_{02},k_{03},k_{04})
\bar{\delta }_k \delta _\omega\  \label{s-1d}\nonumber \eea
where
\begin{eqnarray}
\int du&=&\int_{p_{1}=-\Lambda}^{p_{1}=\Lambda}\frac{dp_{\alpha
,1}}{2\pi}
\int\frac{d\omega_{1}}{2\pi}\int_{p_{2}=-\Lambda}^{p_{2}=\Lambda}\frac{dp_{\alpha
,2}^{(i)}}{2\pi}
\int\frac{d\omega_{2}}{2\pi}\dots\int_{p_{4}=-\Lambda}^{p_{4}=\Lambda}\frac{dp_{\alpha
,4}^{(i)}}{2\pi} \int\frac{d\omega_{4}}{2\pi}\nonumber\\
\delta _k &=&\delta[\nu_{i1} (k_{01}+p_{1})+\nu_{i2}(k_{02}+p_{2})-\nu_{i3}(k_{03}+
p_{3})-\nu_{i4}(k_{04}+p_{4})) \ ,
\end{eqnarray} 
with $\nu_i=\pm $ when $i=R,L$ and $\delta_\omega =\delta(\omega _1+\omega _2+\omega _3+\omega_4)$.
It is not difficult to show ( for example, in a real space nearest-neighbor model, for details 
see  \cite{Shankar}) that the interaction momenta-dependent couplings  are of the form
\beq
V_+ (k_{1,}k_{2,}k_{3},k_{4})=U_{0}\sin(\frac{k_{1}-k_{2}}{2})
\sin(\frac{k_{3}-k_{4}}{2})\cos\left(\frac{k_{1}+k_{2}-k_{3}-k_{4}}{2}\right)
\eeq
with $\mathbf{k}=\mathbf{k_0}+\mathbf{p}$, and where $U_0$ is an ultra local (on-site) repulsive Coulomb 
potential. Two main possibilities exist for the scaling behavior of the coupling functions in (\ref{s-1d}). 
If  $k_{i}$ and $k_{j}$ arise from opposite points in the Fermi surface,
\beq 
V\sim \sin(\frac{k_{1}-k_{2}}{2})\sim \sin(\pi /2+O(p)) 
\eeq
On the other hand, if $k_{i}$ and $k_{j}$ arise from the same side of the Fermi surface 
(Umklapp process)
\beq
V_{Umklapp}=(p_{1}-p_{2})(p_{3}-p_{4})\sim O(p^{2})
\eeq
We are now ready to study the behavior of each term in the effective action 
for small energies and momenta (relative to the Fermi surface) using the
RG techniques. The change in scale is given by:
\begin{eqnarray}
\omega '=s\omega \nonumber\\
p'=sp \nonumber\\
 x'=s^{-1}x
\end{eqnarray}
where $s$ is a small parameter such that $s\rightarrow 0$.
The fields scale according to:
\beq \psi '(\omega '_{i,}p'_{i})=s^{-3/2}\psi (\omega
_{i,}p_{i})\ ,
\eeq
so that the terms linear in $p$ (and the frequency terms) are marginal, while the terms 
$O(p^2)$ are irrelevant.

The scaling of the interaction terms (which are free of $k_0$)
can be deduced from the scaling properties of the delta-functions:
\beq 
\bar{\delta} (p)\rightarrow \bar{\delta} (p'/s)=s\bar{\delta}(p')
\eeq
\beq \delta (\omega)\rightarrow \delta (\omega'/s)=s\delta
(\omega')]\eeq
Then the inter-valley  processes (those in which $k_{i}$ and $k_{j}$
come from different Fermi points) are marginal. The
Umklapp process are suppressed by a term $0(p^2)$,{\it i.e.} they scale as
$s^2$ and are, therefore, irrelevant. The scaling of the gauge
potential depends on the explicit form of $A_0$. For shallow
impurity donors, it scales as $\sim 1/|r-r'|$ (Hidrogen-like 
Coulomb potential) so that it goes as $s^3$ in three
dimensions, and as $s^2$ in two. Hence, they are also
irrelevant\footnote{Ultralocal potentials of the form
$\partial\delta(x-x')/\partial x$ that also break parity
are be marginal.}.

Hence, the complete EFT is given by the following action:
\bea S &=&\sum _{i,\alpha}\int _{-\Lambda }^{\Lambda }
\frac{dp_{\alpha}}{2\pi}\
\frac{d\omega_i}{2\pi } \left\{ \psi_i ^{*}(\omega _{i},p_{i})
\left[i\omega -\lambda p_{i}\right]\psi_i (\omega _{i,}p_{i})\right\}\nonumber\\
&&+\frac{1}{4}\sum _{i}\int d\mu
\psi_{i_1} ^{\dagger }(1)\psi_{i_2} ^{\dagger }(2) V \psi_{i_3}(3) \psi_{i_4}(4) 2\pi \bar{\delta }_p\delta _\omega\\
\label{eftaction}
\nonumber
\eea
Here $V$ is a coupling constant (Umklapp processes are excluded).
The simple picture that emerges is displayed in figure (\ref{Pockets}): the 
low-energy dynamics of the three-dimensional Fermi surface
can be described in terms of six one-dimensional 
Luttinger systems.

\begin{figure}[ht]
\centering \leavevmode
  \epsfxsize 3.2 in
  \epsfbox{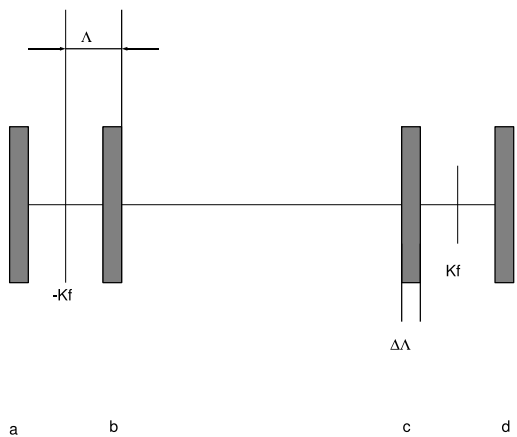}
   \caption{Regions of momentum space around each Fermi point, defined by a, b, c,  d,  being integrated out. They lie a distance $\lambda$ from the Fermi points L and R and have a with $\Delta \Lambda$ \label{cap:integrated-out}}
\end{figure}
\begin{figure}[ht]
\centering \leavevmode
  \epsfxsize 2.0in
   \epsfbox{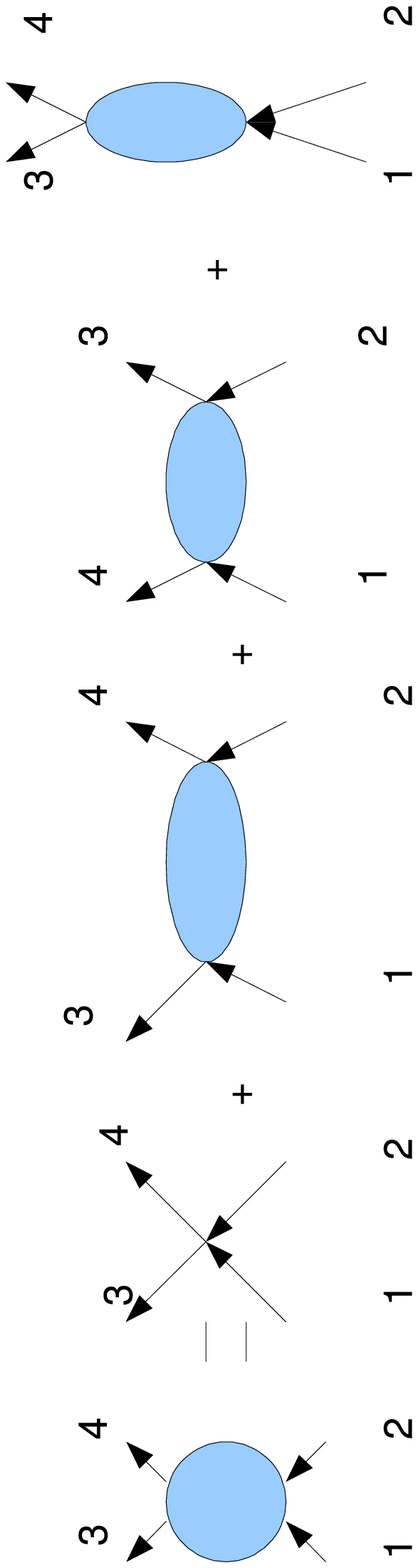}
   \caption{Feynman Diagrams at second order in the coupling constants}
\label{Feynman-Diagrams}
\end{figure}

Since the quartic terms in the EFT action (\ref{eftaction}) are marginal, it is 
necessary to consider their quantum corrections, {\it i.e.}, to include one-loop contributions in the 
RG calculations.
This is done by writing down all the Feynman diagrams up to second order in the interaction coupling
constants and integrating out all momenta in a shell of width $\Delta \Lambda$. 
For one-dimensional systems, this integration is made about the four Fermi lines 
(labeled $a,b,c$ and $d$ in Fig(\ref{cap:integrated-out})) 
located at distance $\Lambda$ from the Fermi points
with width $\Delta \Lambda$.
In the regime of small momenta, the Coulomb potential is usually taken to be
constant,
and all the external frequencies and momenta are set to zero, obtaining
\begin{eqnarray}
 \Delta V &=&\frac{V_{o}^{2}}{4\pi ^{2}}
 \int d\omega \int _{\Delta \Lambda }dp
\frac{1}{\left[i\omega -E(p)\right]\left[i\omega -E(p)\right]}\nonumber\\
&&- \frac{V_{o}^{2}}{4\pi ^{2}}
 \int d\omega \int _{\Delta \Lambda }dp
\frac{1}{\left[i\omega -E(p)\right]\left[i\omega +E(p)\right]}\nonumber\\
&&- \frac{V_{o}^{2}}{4\pi ^{2}}
 \frac{1}{2}\int d\omega \int
_{\Delta \Lambda }dp\frac{1}{\left[i\omega
-E(p)\right]\left[-i\omega -E(p)\right]}
\end{eqnarray}
where we have used the symmetries $E(p)=E(-p)$ and $E(p\pm \pi )=E(p)$
(see figure (\ref{Feynman-Diagrams}) ).

Evaluation of the integrations leads to the Beta function 
$\beta= \Lambda dV/d\Lambda$:
\beq 
\beta (V)=\left[\frac{V^{2}}{2\pi }-\frac{V^{2}}{2\pi
}\right]=0
\eeq
This result implies the existence of a line of continuous fixed points ( see, {\it e.g.} \cite{Shankar}). 
The emerging scale invariant states are identified with those of a Luttinger liquid. 
It follows from the above calculation that the Luttinger liquid phase is stable against 
quantum corrections.
This perturbative analysis makes sense if the low-energy effective
dynamics picture of figure (\ref{Pockets}) is stable under perturbations, which is 
a hypothesis we have implicitly made.

\section{Strong coupling regime and the Mott transition }

As we have discussed in the previous section, the EFT of the
three-dimensional crystalline structure of $Si$-doped $GaAs$ 
is given by a system of six coupled one-dimensional  
Luttinger liquids, which include Umkplapp processes. 
In order to gain some insight about the behavior of the system,
we will neglect for the moment the coupling among the Luttinger liquids, 
so that we can focus on reduced one-dimensional dynamics.
Each Luttinger system is exactly solved \cite{Yang-Yang}
and from their exact solution, it is known  
that the excitations of the system include charge density waves.
From the RG point of
view, as one moves along the line of fixed points, the scaling
dimensions of the various fields change continuously away from the 
free field fixed point values. Simultaneously, the Umklapp processes
that were initially suppressed by a factor $O(p^2)$ change their scaling behavior, 
firstly becoming marginal and later on, relevant operators.
In this new scenario,  their effects should be also considered in the low-energy 
description of the system.

In generic conditions, the Umklapp process are short-distance,
high-energy effects which are not considered in the EFT, since the
scale of the involved momenta is of the order of $\pi/a$ ($a$ is the
microscopic lattice spacing).
In fact, the low-energy physics depends on short distance
dynamics only through relevant and marginal couplings, and 
possibly on a few irrelevant terms, if one measures small enough 
effects \cite{Polchinski}. However, 
for the class of systems we are considering in this paper,
the low disorder impurities extend the microscopic lattice 
symmetries to a mesoscopic domain,
implying that Umklapp processes are also promoted to
mesoscopic scales. 
The transferred momentum between lattice and electrons 
at these scales is now of the order of $\pi/a^*$,  where $a^*$ is identified 
in the physical system with the effective Bohr radius of the
impurities ($a^* \sim  100 A $). Note that this mechanism is not
fine tuned since $ a^*$ depends on the screening and therefore,
on the density. There will be a range of values of $ a^* $,
given by a corresponding range of the doping density $n$, for which 
Umklapp processes should be included in the low energy theory.
The appearance of Umklapp process in this systems is also interesting 
because they have been put in correspondence with the Mott transition. 
Their effects can be naturally included in the EFT 
after bosonization of each Luttinger liquid.

One dimensional systems may be described in terms  of
density and spin fluctuations (see, {\it e.g.}, \cite{Giamarchi-2}
and \cite{Giamarchi-3})
\beq
H\ =\ H_{\rho }\ +\ H_{\sigma }\ +\ H_{scatter}\ +\ H_{Umklapp}
\eeq
with
\begin{eqnarray}
&& H_{\nu }=\frac{1}{2\pi }\int dx\left[u_{\nu }K_{\nu }(\pi \Pi
_{\nu })^{2}+\frac{u_{\nu }}{K_{\nu }}\left[\partial _{x}\phi
\right]^{2}\right]\label{1d-bosonized} \\
&& H_{Umklapp}=\frac{2g_{3}}{(2\pi a|^{2}}\int dx
\cos[\sqrt{8}\phi _{p}+\delta x]\label{1d-bosonized-2} \\
&& H_{scatter}=\frac{2g_{3}}{(2\pi a|^{2}}\int dx
\cos[\sqrt{8}\phi _{\sigma }]\ .\label{1d-bosonized-3} \\ \nonumber
\end{eqnarray}
Here $\nu =\rho $ or $\sigma $ denote the charge and spin labels, and the bosonic fields 
$\phi _{p}$ and $\phi _{\sigma }$ describe the 
charge and spin fluctuations, respectively.
The parameters $K_{\rho}$, $K_{\sigma}$, $u_{\rho}$, $u_{\sigma}$ and
$g_3$ are the only necessary data to describe the long distance
physics. The parameter $\delta =4k_{F}-2\pi /a$ is measure of the distance 
away from half filling, and $\pi /2a$ is the Fermi vector at half filling.
Therefore, the doping $d$ is related to $\delta$ through $d=a\delta/2\pi$.
The RG equations in these variables yield the system:
\begin{eqnarray}
 &&\frac{dK}{dl}=-\frac{1}{2}y_{3}^{2}K^{2}J_{0}(\delta (l)a) \\
 &&\frac{dy_{3}}{dl}=(2-2K)y_{3}\\
 &&\frac{du}{dl}=-\frac{1}{2}y_{3}^{2}uKJ_{2}(\delta (l)a)\\
  &&\frac{d\delta }{dl}=\delta (l)+\frac{1}{2\pi a}y_{3}^{2}J_{1}(\delta(l)a) \ ,\\\nonumber
\end{eqnarray}
where $y_{3}= g_{3} /(\pi u)$ and $l$ describe the renormalization of the cutoff $a(l)=a \exp{l}$, and $J_\nu$ are Bessel functions. 
For half filling, the RG equations coincide with those of Kosterlitz-Thouless, 
with a separatrix at $K-1=g_3/(2\pi u)$  between a regime for which $g_3$ is irrelevant and another one for which it becomes relevant.
Given that vortex condensation or limiting surface
creation are known to arise in systems for which the Kosterlitz-Thouless transition
develops, one might expect that these physical processes could similarly appear 
in our case.

The Mott (or more properly, the Hubbard-Mott) transition  is a strongly
correlated effect yielding a metal-insulator transition when
the electron interactions  become strong enough.
Very frequently, it is studied within the realm of the Hubbard model.
\beq H=-\sum _{\langle i,j\rangle ,\sigma }t(c_{i\sigma }^{\dagger
}c_{j\sigma }+c_{j\sigma }^{\dagger }c_{i\sigma })+U\sum_i
n_{i\uparrow }n_{i\downarrow }\ ,\eeq
where {\it t} is the hopping parameter and {\it U} the on-site Coulomb repulsion.
The transition itself is interpreted as arising from the competition between
the kinetic term, which favors the delocalization of electrons and
the interaction (repulsive) term, which favors localization. Since equations 
[\ref{1d-bosonized}],[\ref{1d-bosonized-2}],[\ref{1d-bosonized-3}],
are the most general for one-dimensional fermionic systems, they describe the
Hubbard model in one dimension as well. If fact  for weak coupling, it is
possible to show that:
\beq 
u_{\rho }K_{\rho }=u_{\sigma }K_{\sigma }=v_{F}\qquad u_{\rho
}/K_{\rho }=v_{F}+U/\pi \eeq

\beq u_{\sigma }/K_{\sigma }=v_{F}-U/\pi \qquad g_{3}=U\ .
\eeq
Therefore, the interaction driving the Umkplapp processes is the on-site Coulomb repulsion.
Let us consider the charge sector of the one dimensional theory 
near half filling. Parametrizing the deviation from this
point through the chemical potential we can write the
Hamiltonian as:
\begin{eqnarray}
 H=\frac{1}{2\pi }\int dx\left[u_{\rho}K_{\rho }(\pi \Pi _{\rho })^{2}+
\frac{u_{\rho }}{K_{\rho }}\left[\partial _{x}\phi \right]^{2}+ 
\frac{2g_3}{2\pi\a^2} Cos[\sqrt{8}\phi _{\rho}]-\mu \partial_x \phi_{\rho} \right]
\end{eqnarray}
After introduction of the new field, $\phi=\sqrt{2}\phi_{\rho}$ 
it is possible to rewrite the Hamiltonian (for details see  \cite{Giamarchi-2}, \cite{Luther-Emery-Back})
as:
\begin{eqnarray}
&&H=\sum_k vk(c^{\dagger}_{L,k}c_{L,k}-c^{\dagger}_{L,k}c_{L,k})+
\frac{g_3}{\pi a^2}\sum_k c^{\dagger}_{L,k}c_{R,k}+h.c \nonumber\\
&& + \frac{\pi u Sh(2\theta)}{L}\sum_p 2\rho_+(p)\rho_-(-p)-f_1(\rho_+(p)\rho_+(-p)+\rho_+(p)\rho_+(-p))
\end{eqnarray}
where $\rho_s(p)$ denotes the Fourier transform of the fermionic density ($s=\pm$),
$v=u(\cosh2\theta +f_1 \sinh 2\theta) $
and $\exp(2\theta)=1/(2K_{\rho})$.
For the value $K_{\rho}=1/2$, the so-called Luther-Emery line, the two fermion species 
become non-interacting and the Hamiltonian can be diagonalized by a Bogoliubov transformation, 
given raise to two bands in the spectrum.
\beq
E_{L,R}=\pm \sqrt{(vk)^2+(g_3/(2\pi\alpha))^2}
\eeq
Therefore, the gap energy is given basically by the the Umklapp coupling constant. Away from this line, 
the interaction term gives rise to 
a renormalization of the gap. The energy values are:
\beq
E_{L,R}=\pm \sqrt{(vk)^2+(\Delta/2)^2}\ ,
\eeq
where $\Delta$ is the renormalized gap.
For enough small doping,(near half filling), the chemical potential is located in the gap region and at zero (or low  )temperature, the system becomes insulating. As soon as the chemical potential crosses the upper (or lower) band, a chiral current is established.  This is a brief account of the Mott transition in one dimension.
In two dimensions, further difficulties arise. However, several techniques have been
developed to deal with strongly coupled fields for these systems. The analysis of
the non-trivial fixed-points of the RG in this case is beyond the scope of the
present paper \cite{future}.

The preceding analysis allows us to discuss the order parameter for the Mott transition.
As it has been already discussed by different
authors ( \cite{Georges-Mott-Universality},
\cite{Kotliar-Rozenberg-orderP}, \cite{Castellani}), one of the main
problems in characterizing the Mott transition is the absence of any
obvious order parameter. For the  Hubbard model, Castellani et al have identified it with 
the number of doubly occupied sites, by mapping the system onto the 
Blume-Emery-Griffits model, which describes the mixed phases of $^3He-^4He$.
Moreover, in the context of Dynamical Mean Field Theory, which maps the system onto a 
self-consistent impurity Anderson model, 
it has been argued that the order parameter is related to the effective hybridization of the medium, 
which captures the localizing to delocalizing character of the transition.
For the extended Hubbard model, other possibilities have been discussed in the literature. 
Let us consider here the $d$-density wave order parameter, 
as defined by Wilczeck and Nayak in any number of dimensions:
for a class of systems,
it is assumed the existence of the non-trivial correlation
function:
\beq \langle c_{k+Q,\sigma }^{\dagger }c_{k}\rangle\  =\ if(k)\ ,
\eeq
where the vector $Q$ has components of magnitude $\pi/a$ in every direction 
and $f(k)$ is a real function that changes sign under rotations of $\pi/2$ around any axis. 
This correlation function breaks a number of symmetries, namely time inversion
$T$, $\pi/2$ rotations, and lattice translations 
by one lattice unit, but preserves the product or the square of two of them. As it was pointed out in ref 
\cite{Wilzeck-Nayak1}, this 
correlation function can be considered as arising from an order parameter which halves the size of the Brillouin 
zone and naturally induces the Mott transition. A quantum state realizing this correlation may be 
thought of as a $d$-wave spin singlet. In fact, the the following BCS-like wave function
\beq \Psi=\prod_k (u_{k\uparrow}+c^{\dagger}_{k\uparrow}+ v_{k
\uparrow}C^{\dagger}_{k+Q\uparrow})(u_{k\downarrow}c^{\dagger}_{k\downarrow}+v_{k\downarrow}C^{\dagger}_{k
\downarrow}|0\rangle \eeq
is energetically favorable in the extended Hubbard model.
The relevance of this order parameter for the EFT is as follows: on the one hand,
we expect parity to be broken in the EFT if the $d$-density wave order parameter acquires a non-vanishing
expectation value at (or near to) the Mott transition, as a consequence of the CPT theorem and
charge conjugation symmetry.
On the other hand, it is clear from its definition that the order parameter induces a short (or intermediate) 
scale charge ordering, so that, at the corresponding energy scales, a mass gap in the spectrum 
may develop, leading to an unnatural  EFT. In this case, one is led to expect new physics to arise, 
leading to an EFT  that would look very different from the initial one, implying that a search for a new theory 
should be undertaken, with new degrees of freedom to correctly describe the system\footnote{
In fact, in two dimensions the 
$d$-density wave has been associated to the staggered flux phase \cite{Nayak-orderings}, 
which suggest the coupling of different Luttinger liquids, such that the 
main characteristics of each one ({\it e.g.},the gap formation and the Chiral symmetry breaking) 
are preserved.}.                   

As it has been remarked in \cite{Gorkov}, long spin lifetimes in 
$Si$-$Ga$ $As$ samples appear at densities values near those that 
yield the Mott transition, which suggests 
that the latter might play a role in the anomalous spin behavior. 
At first glance, 
this appears far from obvious, because the Mott transition seems to be a 
phenomenon associated to the charge sector of the theory only.
However, the EFT can give us some information given that we can
compare the relative weight in the Hamiltonian of both 
the Dessehaus and the Coulomb terms at doping levels near 
criticality. Given that we know the scaling of both interaction
terms, we can argue that both should be considered as equally
important since at the strong fixed they become marginal simultaneously 
(since both scale to zero  
as $ s^2$ when the energy scales down as $s$). This RG constraint 
connects both charge and spin sectors of the theory and,
therefore, does not exclude the possibility that an spin
effect, such as the one observed in the anomalous spin lifetimes,
could be related to a Coulomb effect in the strong coupling
regime.

\section{Conclusions}

In this article we have presented EFTs for electrons in crystalline systems 
and have studied the
effect of quenched impurities (low disorder effects) in them. Based on experimental 
evidence and consistency and plausibility arguments, we have considered  
the situation in which impurities form arrays in mesoscopic regions for which 
some or all of the discrete symmetries of the microscopic lattice are preserved
in the long distance limit.
We have also discussed EFTs 
for generic systems with substitutional impurities,
although we did not get into details as to how to deal 
with a general case.
We believe that the EFT point of view is a good complement to
the one based on microscopic models for studying strongly correlated 
electron systems in crystals, providing support to them and 
bringing in a useful bridge between strongly correlated problems and perturbative 
band theory calculations. 
We have obtained the EFT for $Si$-doped 
$GaAs$ systems, obtaining  
from the symmetries of the effective fields all possible single
particle terms in the effective Hamiltonian, which reproduce the known terms 
of the pseudo-potential calculations of band structure approaches and 
the so called Dyakonov-Perel Hamiltonian for the spin-orbit coupling.
Remarkably, the resulting low-energy effective theory is described in terms 
of three (six chiral) one-dimensional Luttinger liquid systems and their corresponding intervalley 
transitions.

\clearpage

\vspace {2 cm}

\def\NP{{\it Nucl. Phys.\ }}

 \def\PRL{{\it Phys. Rev. Lett.\ }}

 \def\PL{{\it Phys. Lett.\ }}

 \def\PR{{\it Phys. Rev.  \ }}

 \def\PRB{{\it Phys. Rev. B  \ }}

 \def\CMP{{\it Comm. Math. Phys.\ }}

 \def\IJMP{{\it Int. J. Mod. Phys.\ }}

 \def\MPL{{\it Mod. Phys. Lett.\ }}

 \def\RMP{{\it Rev. Mod. Phys.\ }}

 \def\AP{{\it Ann. Phys.\ }} .

\end{document}